# Reduction of the bulk modulus with polydispersity in non-cohesive granular solids

Juan C. Petit[1] and Ernesto Medina[2,1]

[1]*Laboratorio de Física Estadística de Sistemas Desordenados, Centro de Física,
Instituto Venezolano de Investigaciones Científicas (IVIC), Apartado 21827, Caracas 1020 A, Venezuela*
[2]*Yachay Tech, School of Physical Sciences & Nanotechnology, 100119-Urcuquí, Ecuador*
(Dated: February 15, 2018)

We study the effect of grain polydispersity on the bulk modulus in non-cohesive two dimensional granular solids. Molecular dynamics simulations in two dimensions are used to describe polydisperse samples that reach a stationary limit after a number of hysteresis cycles. For stationary samples, we obtain that the packing with the highest polydispersity has the lowest bulk modulus. We compute the correlation between normal and tangential forces with grain size using the concept of *branch vector/contact length*. Classifying the contact lengths and forces by their size compared to the average length and average force respectively, we find that strong normal and tangential forces are carried by large contact lengths, generally composed of at least one large grain. This behavior is more dominant as polydispersity increases, making force networks more anisotropic and removing the support, from small grains, in the loading direction thus reducing the bulk modulus of the granular pack. Our results for two dimensions describe qualitatively the results of three dimensional experiments.

## I. INTRODUCTION

The study of the mechanical behavior of granular matter is important since these materials are ubiquitous in nature and are widely used in industrial processes [1–6]. In general, granular materials are composed by a size distribution or polydispersity which strongly affects their macroscopic response. Civil, structural and mechanical engineers use polydispersity to design concrete beams more resistant to external loads [3–6]. Such resistance is achieved by reaching for the maximum packing density of the system, as it is done for high performance concrete [7, 8] and ceramics sintering [9–11]. Higher densities are also correlated with less development of micro-cracks in the system [12, 13]. To avoid such micro-cracks, different grain sizes (such as gravel, sand, ordinary cement, limestone filler and silica fume) are mixed in order to increase packing density. The grain size distribution is also important to characterize the structure of a cataclastic fault material [14], which can be related to its deformation history and mechanical stability. Furthermore, comparison of wave propagation in monodisperse versus slightly ordered polydisperse packings have shown that the speed amplitude of sound waves in the latter is reduced [15]. This is due to contact disorder where dispersive effects are induced.

The effect of polydispersity on force fabric variables in granular packings have also been investigated, where mean coordination, porosity and grain mobilization change when the degree of polydispersity varies [16–18]. Furthermore, the force distribution is broader as the grain size span increases since a large population of grains support forces less than the average[18–20]. These changes make the packing exhibit different macroscopic behaviors when external loads are applied [16, 17, 21]. Effective properties such as bulk modulus, shear modulus and bulk density depend intrinsically on the structure of the packing, which is related to the contact network and the force propagation[17, 18, 22–25]. However, surprisingly the macroscopic friction is not affected by the degree of polydispersity. This independence was demonstrated in refs.[20, 26] due to a compensation between fabric and force anisotropies inside the packing.

Results reported previously, have shown that increasing polydispersity of a compacted granular system reduces slightly the bulk modulus [17, 21]. A preliminary work [27] showed experimentally that the strongest force chains emerge at peak effective stiffness, evidencing a latent relation between both. Despite the results of these works, the effect of grain size on force networks and its contribution to the bulk modulus of a polydisperse packing has not been widely studied.

In this work, we study the effect of polydispersity on the bulk modulus and force network of a stationary packing structure achieved after a number of loading-unloading cycles. In this stationary packing state, the grains develop such an overlap that they cannot move appreciably relative to each other during subsequent loads, a situation better described as a unconsolidated *granular solid*. One of its most salient features is the frustration of rotations at length scales from one grain to clusters of a few grains[25]. In section II, we describe our Molecular Dynamic simulations used to model uni-axial loading-unloading cycles applied to each granular packing. In section III we discuss our results for the bulk modulus as a function of the degree of polydispersity and particle friction. We find that in the stationary state, the bulk modulus decreases with polydispersity. In section IV, we address a possible explanation for the obtained values of the bulk modulus in terms of the force networks and grain size as a function of polydispersity. We characterize the size of the grains at contacts by using the concept of the *branch vector length* $\ell$ between pairs of grains. We find grain sets that support different relations between normal forces and $\ell$. As polydispersity increases, the small grains are increasingly isolated from supporting loading forces, thus decreasing the overall bulk modulus of the macroscopic system. Furthermore, while large grain networks support vertical forces, small grain between predominantly support horizontal forces. We end with a summary and conclusions.

## II. SIMULATION PROCEDURE

The simulation performed here consist of a granular packing composed of 1000 circular grains whose radii are chosen from a uniform distribution in the range of $R \in [R_{\text{av}} - \sigma, R_{\text{av}} + \sigma]$, where $R_{\text{av}} = 0.02$ m is the mean radius and $\sigma$ is the distribution width calculated by $\sigma = R_{\text{av}}\delta$. The degree of polydispersity is varied in the range of $\delta \in [0, 5, 10, 20, 30, 40, 50, 60, 70]\%$, where each value represents one packing. These packings have the sufficient number particles, to be representative of the polydisperse packing. The criteria for the latter assertion have been reported in ref. [28], where they find that uniformly distributed polydisperse packings are statistically well described by simulations of e.g. 1000 particles, robustly above $\delta = 20\%$. All the results in this work discuss polydispersities above 30%[29].

The interaction between a pair of grains is modeled using the linear spring-dashpot contact model, where normal, tangential Hookean springs and damping coefficients are considered. Here normal and tangential stiffness are constant parameters, in contrast from those of the Hertz model, where they depend on th Young modulus and Poisson ratio of the material [25, 30] as well as particle interpenetration [31]. Grain parameters correspond to quartz grains, listed in Table I, and considered previously in ref.[25, 32–35]. We considered a simulation box with periodic boundary conditions in the horizontal direction to avoid wall effects. The box dimensions are $W = 50R_{\text{av}}$ in width and $H = 150R_{\text{av}}$ in height. Gravity is not considered in the simulations since it induces irrelevant stress distributions as compared to the intergranular forces contemplated in the simulation.

The granular packing is constructed by randomly positioning grains inside the box without overlapping and initially fixing intergrain friction at $\mu = 0$. Then, both horizontal walls are move towards the center of the box to compact the system. The walls stop moving when the packing porosity falls below of $\psi = 15\%$. This procedure leads to isotropic packings for $\delta \geq 20\%$, while for $\delta < 20\%$ the distribution is notably crystalline with highly prefered contact angles. Similar procedures have previously been used, where an initially "gaseous state" of the grains are compressed to achieved a dense system and an isotropic network [36–38].

We have shown in a recent work [25] that a stationary state can be found if the packing is subjected to a sufficient number of full compression-decompression cycles. The last cycle defines the *limit cycle*, point at which one reaches a stationary hysteresis loop, i.e, an unchanging route in strain-stress space that closes on itself re-producibly. A detailed information about this cycling procedure is given in [25, 34]. In our simulations, the maximum deformation imposed on the packing is here set to $\epsilon_{yy}^{\max} = 0.1$. At the limit cycle, one reaches a stationary packing where properties are stable under further uniaxial compression. Such a state results in interpenetrations that are above the typical threshold of 1% used in loose granular simulations. Thus, one can assume the stationary packing state as a solid-like granular system or a *granular solid*. This regime is relevant physically as reported in ref.[39], where authors compared simulations of highly compacted granular system with experimental results for jammed packings, obtaining average penetration appreciably above 1%. It is this granular-solid state, we are interested in studying the bulk modulus and force networks as a function of polydispersity and particle friction.

TABLE I. Parameters used in simulation. Values correspond to quartz grains (see [32]), which are frequently found in sedimentary rocks.

| Prop. | Symbol | Value |
|---|---|---|
| Density | $\rho_g$ | 2.65 g/cm$^3$ |
| Normal stiffness | $k_n$ | 191.30 GPa |
| Tangential stiffness | $k_s$ | 183.32 GPa |
| Poisson ratio | $\nu$ | 0.08 |
| Damping coeff. | $\gamma_{n,s}$ | $2 \times 10^{-6}$ g/(cm·s) |
| Micro friction | $\mu$ | 0.1,0.3,0.5 |
| Polydispersity | $\delta$ | [0-70]% |

## III. EFFECT OF POLYDISPERSITY ON BULK MODULUS

Once the limit hysteresis loop is found, after a sufficient number uniaxial loading-unloading cycles applied to the granular pack, the bulk modulus for each packing was calculated using the following expression

$$K = \frac{\Delta\sigma_{yy} + 2\Delta\sigma_{xx}}{3\Delta\epsilon_{yy}}, \quad (1)$$

where a variation of the vertical strain, $\Delta\epsilon_{yy}$, is imposed when monitoring the variation of vertical stress, $\Delta\sigma_{yy}$, and horizontal stress $\Delta\sigma_{xx}$. Eq.(1) is strictly appropriate for macroscopically isotropic systems, and measures how the granular pack responds to changes in the volume in that particular direction. The bulk modulus for each packing increases with vertical stress following a power law of the form $K = K_0 \sigma_{yy}^\alpha$, where the exponent changes between $1/2$ or $1/3$ as reported in refs.[25, 34, 40–42]. Previous works have demonstrated that such power law is due to the increase of the mean coordination number during compression, leading to different $\alpha$ exponent [31, 34, 43]. On the other hand, recent works [25, 31] have shown that the degree of polydispersity only weakly changes the $\alpha$ exponent, varying by no more than 4%, but it changes the $K_0$ factor (see ref.[31]). Furthermore, the latter reference also demonstrates that the results for the longitudinal elastic moduli resulting from Hertz and linear models are quite close so the linear model used here is fair approximation to more realistic contact models.

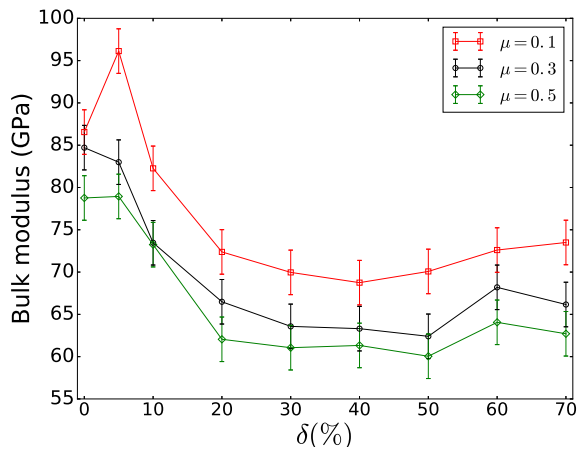

FIG. 1. Bulk modulus as a function of polydispersity for different particle frictions, at the stationary limit. The error bars show that the decrease in bulk modulus as a function of polydispersity is significant. The error bars were obtained by averaging over 10 different packings with the same degree of polydispersity and particle friction.

Figure 1 shows the values of the bulk modulus as a function of polydispersity for different particle frictions. We observe that the bulk modulus decreases with polydispersity and the inter-particle frictions considered. This result is in agreement with those obtained in compressional three-dimensional granular packings [16, 17, 21], where denser packings are achieved since contact deformations and grains rearrangements occur during compression. Furthermore, Fig.2 shows that the mean coordination number decreases, while porosity reaches a maximum, for $\delta = 40\%$, to then decrease weakly as polydispersity and particle friction grow. These results tell us that the grain size affects the bulk modulus by changing mean coordination and porosity. The reduction of the packing fraction with polydispersity is associated with the decrease of the bulk modulus. We think that the stationary pack develops an effective highly porous character due to the distribution of forces in the polydisperse case, where the smaller particles are encaged in pockets whose walls provide support for the external load. Then, the smaller particles are weakly coupled to the supporting structure and thus the bulk modulus is reduced. Another evidence for this is the reduction of the mean coordination number when the degree of polydispersity increases which is also linked to a reduced of the bulk modulus. These results are supported by previous mean field theories where the bulk modulus is proportional to a power law of the mean co-ordination number (see [36, 37, 40, 41]) and has also been demonstrated in recent simulations [31].

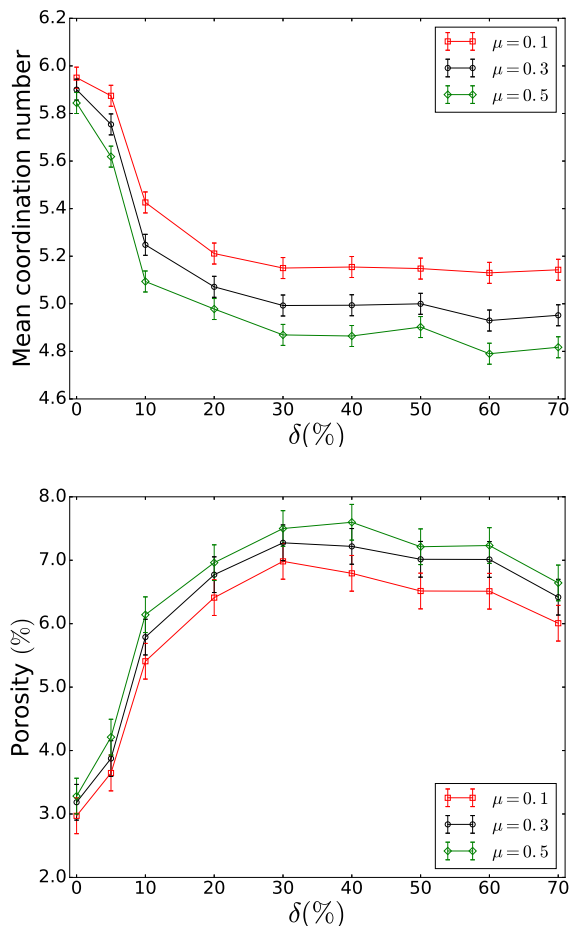

FIG. 2. Top panel: mean coordination number, and Bottom panel: porosity as a function of the degree of polydispersity for different particle frictions. While polydispersity increases mean coordination is reduced and porosity is increased rendering the bulk modulus lower. The effect is enhanced by the local friction.

Previous works in highly polydisperse packings composed of disks [20, 44] and pentagonal grains [26], have shown that strong forces propagate more through larger particles (particle larger than the average particle size) as polydispersity increases. However, it is not well understood how these large and small particles carrying forces contribute to the bulk modulus of a given polydisperse packing. In the next section, we address the effect of the grain size on force networks in order to explore how the force network can be linked to the bulk modulus of the packing.

## IV. EFFECT OF GRAIN SIZE

### A. Force networks

The granular pack forms a contact network through which each contact carries a particular magnitude of the force. The grain size in a contact network can be characterized by using the concept of *branch vector length* $\ell$, as it has been used previously [20, 26, 44]. The branch vector length is defined by the distance between the centers of two grains in contact. This definition allows us to break up the contact network of a polydisperse packing into two parts: i) one denoted by *long contact lengths* ($\ell > 2R_{av}$), where at least one large grain ($R_i > R_{av}$) forms the contact, and ii) denoted by *short contact lengths* ($\ell \leq 2R_{av}$), where only small grains ($R_i \leq R_{av}$) form the contacts. With this in mind, we can relate the grain size with force networks inside the packing as polydispersity changes.

Figure 3 depicts the average magnitude of the normal force $\langle F_n \rangle_\ell$ as a function of $\ell$ for four packings with different polydispersity. This figure depicts the contact lengths carrying strong and weak magnitudes, i.e., above and below the average magnitude of the normal force $\langle F_n \rangle$, respectively. Before compaction, monodisperse packings have all contact lengths equal to $\ell = 2R_{av}$, while for polydisperse packings $\ell$ changes according to size distribution. The monodisperse case exhibits a linear relation between $\langle F_n \rangle_\ell$ and $\ell$, showing only short contact lengths ($\ell < 2R_{av}$), as can be seen in Fig.3. This means that those contacts with $\ell/2R_{av} < 0.9$ exhibit a considerable interpenetration and are able to carry strong forces, while those with $0.9 \leq \ell/2R_{av} < 1.0$ carry the weak normal forces. For packings with $\delta \leq 5\%$, a similar linear relation between $\langle F_n \rangle_\ell$ and $\ell$ is also found. This linear relation is regarded as the effective response of the packing to external forces. The interpretation for the limit case is obvious since the undeformed grains are all of equal radius and deformation (shorter contact lengths) are in direct relation to the normal force applied. As we sample shorter contact lengths, we have higher forces applied and this implies a negative slope for the normal forces versus the contact lengths. The extrapolation of the straight line cuts at zero indicating no force applied to contacts where $\ell = 2R_{av}$ as expected. We can also think of this limit as the affine regime.

When the degree of polydispersity increases, the analysis is more complex since the contact lengths do not map trivially onto deformations. One sees a preservation of the linear relation between applied forces and contact lengths, only at the higher and lower ends of the contact length scale. The same negative slope for $\langle F_n \rangle$ versus $\ell$ as in the monodisperse limit or affine regime.

For intermediate contact lengths and the larger polydispersities, a new macroscopic response is found where smaller contact lengths carry smaller forces while larger contact lengths carry the larger forces. Another linear relation develops (with a slope inversion) that describes how the geometry of the granular solid distribute the applied forces. For $\delta > 30\%$, smaller grains can fit between the space of the larger ones, changing the trend between $\langle F_n \rangle_\ell$ vs $\ell$ (see Fig.3). If this is the case the smaller grains are shielded and short contact lengths carry only lower than average normal forces. This latter case is more pronounced as the degree of polydispersity increases, suggesting the increased participation of large grains to support normal forces. These results support quantitatively, previous results given in refs.[20, 26, 44], where they showed that large grains support strong forces.

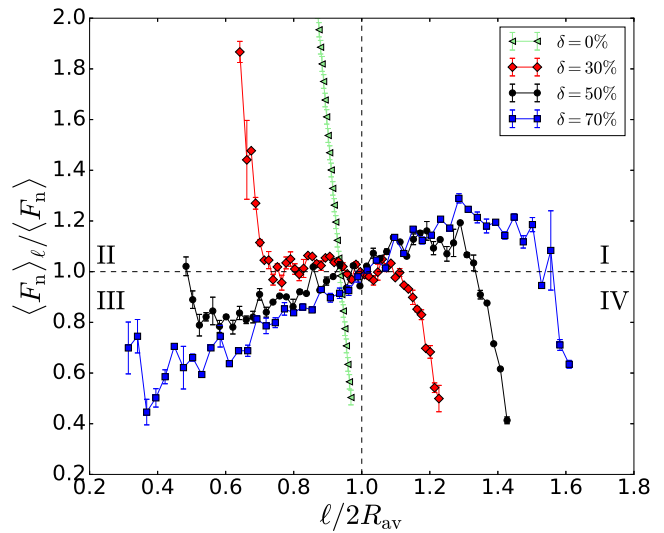

FIG. 3. Average magnitude of normal force for a particular $\ell$ as a function of the branch vector length for four packings with different polydispersity. These data correspond to the final loading state of the final cycle. Dashed lines correspond to the values, $\langle F_n \rangle_\ell = \langle F_n \rangle$ for y axis and $\ell = 2R_{av}$ for x axis. Interparticle friction was set to $\mu = 0.3$.

Figure 4 top panel shows the force network for the extreme case (for clarity) of 70% polydispersity. It depicts those contacts carrying normal forces above (dark lines) and below (light lines) the average. One can see a clear tendency for long contacts to carry above average forces while short contacts carry weaker forces. There is a bimodal character to the network of forces as previously demonstrated in refs.[35, 38]. While one can readily notice a continuos load support for the dark line network, the lighter line network is isolated into disconnected pockets.

Perhaps a clearer picture of the latter observation can be obtained just depicting contact lengths having values above and below the average, disregarding, this time, the force magnitude. This allows us to explore the distribution of contact lengths inside the packing. Fig.4 bottom panel shows more clearly that short contact lengths are concentrated in small clusters isolated by long contact lengths, which represent a unique connected network contributing to the elastic behavior of the packing. A careful comparison of the force and length networks, one can see that those clusters of short contact lengths mostly support weak forces.

The previous results suggest that the connection between polydispersity and the behavior of the bulk modulus is that a non-uniform burden on the grains. They make for an effective porosity as a function of polydispersity that renders the bulk modulus smaller as seen before (see Fig.2). This interpretation is closely related to recent works on stress distributions in porous media, where increasing the degree of pore disorder, i.e. number and size of the pores, the tensile strength and elastic moduli decreases [45, 46]. Such reduction is due to stress concentrations around pore clusters, more pronounced at high porosities than at low porosities. The sponge-like nature of the force support, as one increases the polydispersity, can be thought of as a highly porous structure, resulting in a lower bulk modulus.

One noticeable feature in Fig.3 is that there is a threshold behavior for the slope inversion close to $\delta = 30\%$ where a large range of contact lengths carry the average normal force. This is an interesting feature since the behavior is tantamount to a regular fluid under pressure when one ignores the action of gravity, as in our case.

When polydispersity continues to increase, the range of contact lengths widens emphasizing large grain contacts (larger contact lengths) supporting large forces and small grain contacts carrying small forces. The slope inversion region expands and shows a well defined limit slope above 40% polydispersity.

We revisit the issue particle interpenetration depth as a function of the polydispersity: After several loading-unloading cycles, particle motion is suppressed and particle penetrations dominate during packing compression reaching a stationary state. This leads intuitively to a higher packing fraction and thus a higher bulk modulus. To quantify this, we have estimated the mean interpenetration, $\xi_{\mathrm{av}}$, achieved for each packing with different polydispersity at the limit cycle. Figure 5 shows that the mean interpenetration decreases with polydispersity for a particular value of $\mu$. This is an interesting result since it suggests that the highest value of the bulk modulus obtained for the random monodisperse packing was due to the development of more contacts originated by particle interpenetration during loading. This is the reason why the monodisperse packing has reached a packing fraction above and a mean coordination number close to the hexagonal packing value, $\phi_{\mathrm{Hex}} = 92\%$ and $Z_{\mathrm{Hex}} = 6$, respectively. Increasing polydispersity, the average interpenetration decreases causing a lower mean coordination number and less packing fraction or high porosity (see Fig.2). When particle friction increases, the mean interpenetration increases for a particular $\delta$ due to the frustration of particle slidings during compaction [25]. Interpenetration is the only way to accommodate for the additional force applied.

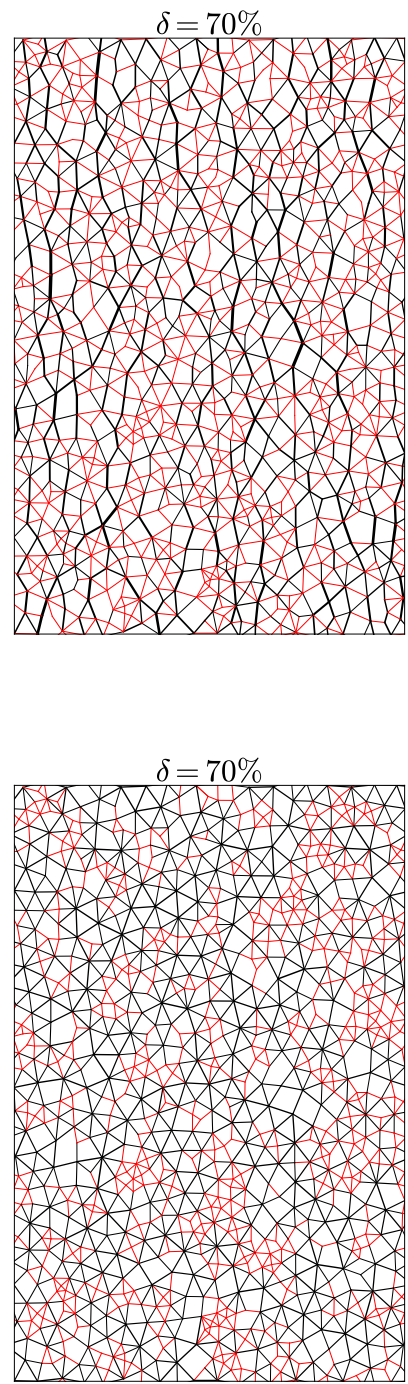

FIG. 4. Top panel: Normal force networks. Light and thin contacts (red online) depict normal forces below average, while dark and thick contacts depict normal forces above the average. Bottom panel: Length networks. Light contact lengths (red online) depict short contact lengths (below than average radii), while dark contact lengths depict long contact lengths (above the average radii). One can readily see how short contact lengths are encaged (do not percolate) by longer contact lengths, that percolate.

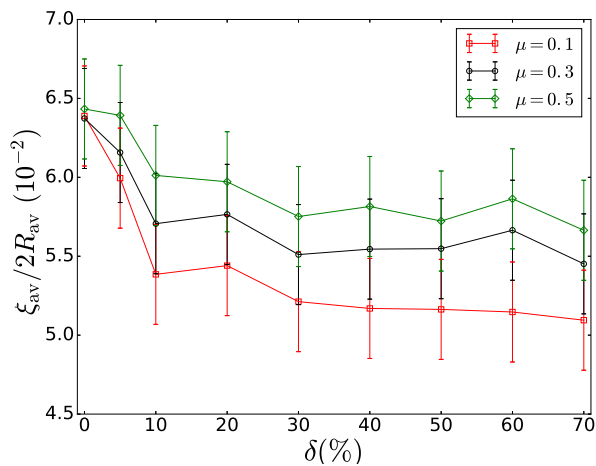

FIG. 5. Mean interpenetration as a function of polydispersity for different particle friction.

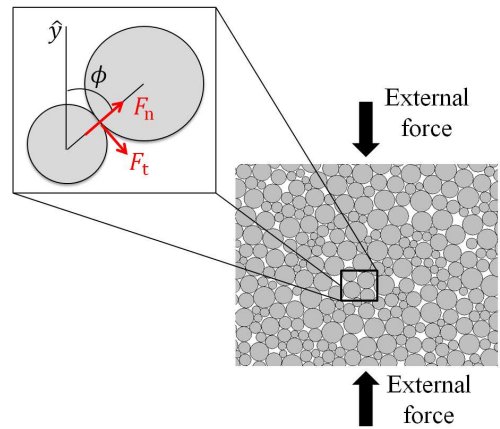

FIG. 6. An illustration of the granular packing subjected to a vertical external force and the contact orientation of two grains with their respective normal and tangential forces evaluated in Figs. 7 and 8.

## B. Orientations and anisotropy of forces

When a loading state is applied on the granular pack, a fraction of the force chains are oriented parallel to the loading axis, while the others are oriented at certain angles $\phi$ thereof. A recent work [27] has shown experimentally, that the correlation between vertical force chains increases with macroscopic load. However, in a polydisperse packing, it is not clear how these strong vertical forces are carried by the granular system.

In order to elucidate this, we have quantified the orientation of the average magnitude of the normal and tangential contact forces, (see Figure 6), focusing contact lengths larger or smaller than the average. The normal force was measured respect to the $y$ axis, while tangential force respect to the $x$ axis. Figure 7 shows the orientation of normal contact forces for three packings with different polydispersity. We have obtained that for long contact lengths, $\ell > 2R_{av}$, strong normal forces are carried by contacts oriented within a range of angles with respect to the loading direction. For $\delta = 70\%$, strong forces are oriented at angles $\phi \in (-50°, 50°)$, while, weak normal forces correspond to those contacts oriented at higher angles. As polydispersity increases from 30-70% the vertical force is increasingly placed on large contact lengths. On the other hand, for short contact lengths, $\ell \leq 2R_{av}$, the strong normal forces concentrate in a smaller range of angles, $\phi \in (-45°, 45°)$, while weak forces concentrate on a wider range.

As the polydispersity increases short contact lengths carry a lower proportion of the vertical forces. These results indicate that those long contact lengths oriented vertically are more predominant carrying strong normal forces as polydispersity increases in qualitative agreement with references [20, 26, 27, 44].

Figure 8 shows the orientation of tangential contact forces for three packings with different polydispersity. For both long and short contact lengths, they exhibit the maximum tangential force oriented at angles close to $\phi = \pm 45°$. This maximum increases with $\delta$ for long contact lengths, while it decreases with short ones. This shows that long contact lengths also contribute to carrying strong tangential forces as polydispersity increases.

The data shown in Fig.7 and Fig.8 can be well described by using general expressions of the form

$$\frac{\langle F_n \rangle_\phi}{\langle F_n \rangle} = m_n + a_n \cos(p_n\{\phi - \phi_n\}), \qquad (2)$$

$$\frac{\langle F_t \rangle_\phi}{\langle F_n \rangle} = a_t \sin(p_t\{\phi - \phi_t\}), \qquad (3)$$

where $m_n$ is a fitting parameters close to one. $p_{n,t} \approx 2$ for $\delta \in [20 - 70]\%$, and $\phi_n$ and $\phi_t$ represent privileged angles, which tend to follow the principal stress direction ($\phi_n = \phi_t = 0°$) for a vertically compacted system. $a_n$ and $a_t$ are positive variables measuring the anisotropy of normal and tangential forces inside the packing. Anisotropy means that the orientation distribution of normal and tangential forces deviates from an uniform distribution. We can see in Fig.8 that $\langle F_t \rangle(\theta)$ has positive and negative values with the same amplitude meaning that each value generates an opposite torque. We have also checked that $\langle F_t \rangle \to 0$ consistent with the orthogonality requirement stated in previous works [47, 48].

From Eq.(2) and Eq.(3) one can determine the anisotropy as a function of $\delta \in [20 - 70]\%$ and particle friction, focusing on long and short contact lengths. For values of $\delta < 20\%$, Eq.(2) and Eq.(3) do not describe well the data because contact orientations are very concentrated. Figure 9 shows that for long contact lengths both normal and tangential anisotropy increase when polydispersity and particle friction increase. For short contact lengths, both anisotropies decrease with $\delta$ but still increase with $\mu$. This means that large grains increase the force anisotropy inside the packing with increasing polydispersity and particle friction.

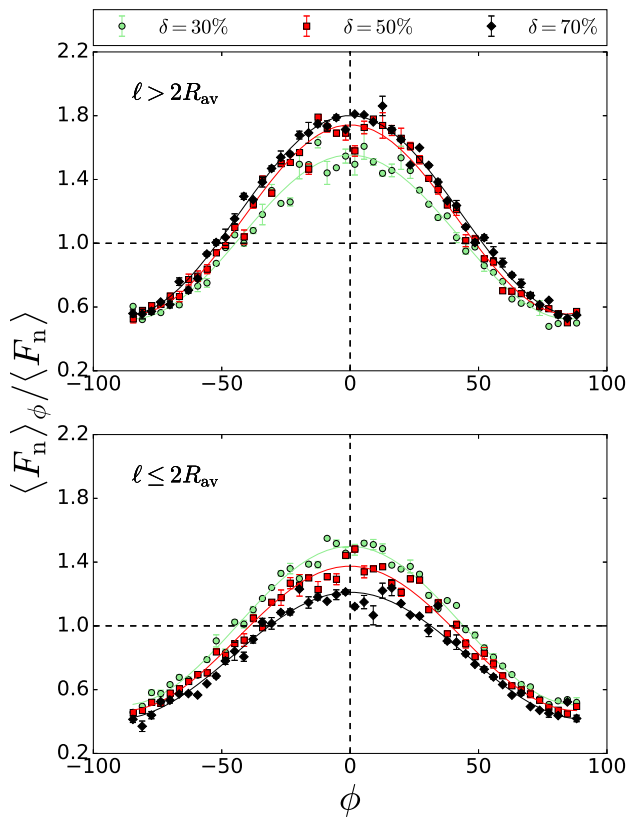

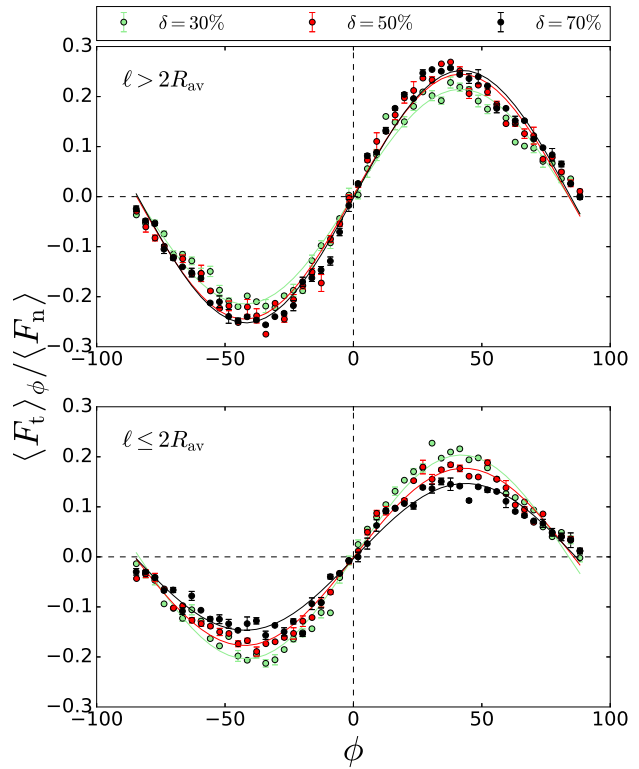

FIG. 7. Orientation of normal contact forces inside the packing for different polydispersity. Top panel: long contact lengths. Bottom pannel: short contact length. Data corresponds to the final loading state of the final cycle. All packings have an inter-particle friction of $\mu = 0.3$.

FIG. 8. Orientation of tangential contact forces inside the packing for different polydispersities. Top panel: long contact lengths. Bottom panel: short contact lengths. All packings have an inter-particle friction of $\mu = 0.3$.

## V. SUMMARY AND CONCLUSIONS

We studied the effects of polydispersity on the bulk modulus and force networks in two dimensional non-cohesive granular solids. The system studied is a cycled granular pack (compression-decompression) that has reached stationary properties under uni-axial stress. We found that the bulk modulus for the stationary pack decreases with polydispersity and particle friction, showing the highest value for the monodisperse packing. In order to shed light on these results, we analyzed the effect of the grain polydispersity on the force networks within the sample. The grain contacts were characterized by the branch vector or contact lengths, which allowed us to break up the contact network into those with contact lengths above and below the average in the pack. We also assessed the forces carried by the contacts and classified them below and above the average contact force. We found that long contact lengths concentrate the largest normal forces and are oriented within a range around the vertical and bear the maximum normal and tangential force more frequently as the degree of polydispersity increases. On the other hand, the small contact lengths are increasingly isolated in cages created by large contact lengths that isolate the smaller grains from the external stress. This caging effect renders the granular solid effectively porous with a concomitant expected reduction of the bulk modulus. The local friction increases the porosity generating mechanism by frustrating particle rearrangements that can lead to higher packing densities. Our results are in qualitative agreement with recent experiments measuring the uniaxial stress distribution in porous media, albeit comparing two a three dimensional systems.

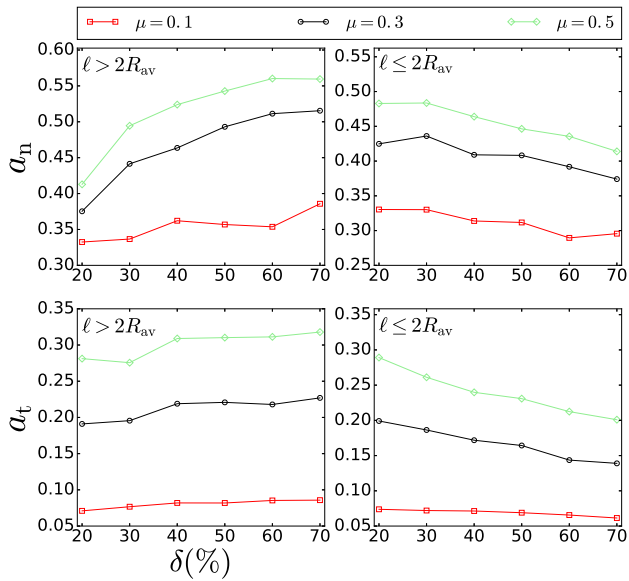

FIG. 9. Fitting parameters according to Eqns. 2 and 3, for normal and tangential force anisotropy as a function of polydispersity for different particle friction.

# ACKNOWLEDGMENTS

We gratefully acknowledge fruitful discussions with Xavier García and Ivan Sánchez.